\documentclass{aa}  
\usepackage{bm}
\usepackage{natbib}
\usepackage{subcaption}
\usepackage{multicol}
\usepackage[mathlines,switch, columnwise]{lineno}
\let\oldequation\equation
\let\oldendequation\endequation
\renewenvironment{equation}
  {\linenomathNonumbers\oldequation}
  {\oldendequation\endlinenomath}

\bibpunct{(}{)}{;}{a}{}{,} 
\newcommand{\vect}[1]{\mathbf{#1}}
\renewcommand{\vec}{\vect}
\newcommand{\td}[2]{\frac{d #1}{d #2}}

\newcommand{\pd}[2]{\frac{\partial#1}{\partial#2}}

\newcommand{\pdd}[2]{\frac{\partial^2#1}{\partial#2^2}}

\usepackage{graphicx}
\usepackage{txfonts}
\begin{document}

\title{Interplanetary spread of  solar energetic protons near a high-speed solar wind stream}
\titlerunning{Solar energetic proton spreading in a high-speed solar wind stream}        
   \author{N. Wijsen
          \inst{1,2}
          \and
          A. Aran\inst{2}
          \and
          J. Pomoell\inst{3}
          \and
          S. Poedts\inst{1}
          }

   \institute{Department of Mathematics, KU Leuven, Belgium\\
              \email{nicolas.wijsen@kuleuven.be}
          \and
                 Departament F\'{i}sica Qu\`antica i Astrof\'{i}sica, Institut de Ci\`encies del Cosmos (ICCUB), Universitat de Barcelona, Spain
         \and
                 Department of Physics, University of Helsinki,  Finland
             }

   \date{Received 26 January 2019 / Accepted 18 February 2019}

% \abstract{}{}{}{}{} 
% 5 {} token are mandatory
 
  \abstract
   {}
   { We study how a fast solar wind stream embedded in a slow solar wind influences the spread of solar energetic protons in interplanetary space. 
   In particular, we aim at understanding how the particle intensity and anisotropy vary along interplanetary magnetic field (IMF) lines that encounter changing solar wind conditions such as the shock waves bounding a corotating interaction region (CIR).
  Moreover, we study how the intensities and anisotropies vary as a function of the longitudinal and latitudinal coordinate, and how the width of the particle intensities evolves with the heliographic radial distance. Furthermore, we study how cross-field diffusion may alter these spatial profiles. %obtained dependencies.
}
   {To model the energetic protons, we used a recently developed particle transport code that computes particle distributions in the heliosphere by solving the focused transport equation (FTE)
in a stochastic manner. 
The particles are propagated in a solar wind containing a CIR, which was generated by the heliospheric model, EUHFORIA. 
We study four cases in which
we assume a delta injection of 4 MeV protons spread uniformly over 
different regions at the inner boundary of the model. 
These source regions have the same size and shape, yet are shifted in longitude from each other, and are therefore  magnetically connected to different solar wind conditions. 
    }
   {
The intensity and anisotropy profiles along selected IMF 
lines vary strongly according to the different solar wind conditions encountered along the field line. 
The IMF lines crossing the shocks bounding the CIR show the formation of accelerated particle populations, with the reverse shock wave being a more efficient accelerator than the forward shock wave. 
The longitudinal intensity profiles near the CIR are highly asymmetric in contrast to the profiles obtained in a nominal solar wind.
For the injection regions that do not cross the transition zone between the fast and slow solar wind, we observe a steep intensity drop of several orders of magnitude near the stream interface (SI) inside the CIR. 
Moreover, we demonstrate that the longitudinal width of the particle intensity distribution  can increase, decrease, or remain constant with heliographic radial distance, reflecting the underlying IMF structure.
 Finally, we show how the deflection of the IMF at the shock waves and the compression of the IMF in the CIR deforms the three-dimensional shape of the particle
distribution in such a way that the original shape of the injection profile is lost. 
   }
   {}


   \maketitle
%
%________________________________________________________________

\section{Introduction}
Understanding and modelling the transport of solar energetic particles (SEPs) in the solar wind remains a major challenge in space physics. Although the propagation and momentum changes of SEPs through the heliosphere may be described
%are fairly well described \alert{how do we actually know this}\textcolor{blue}{good question. To what extent is the test-particle approach of the FTE valid} 
by the focused transport equation (FTE) \citep[see e.g.][]{roelof69,ruffolo95,isenberg97,leRoux09}, the resulting phase-space distribution functions
%describing the energetic particles
are strongly dependent on the assumptions regarding the solar wind and the associated interplanetary magnetic field (IMF). 
On the smallest scales, the importance of the solar wind structure is reflected through the  effect of turbulence on particle transport. 
Turbulent fluctuations can act as scattering centres, changing  both the direction of propagation and the speed of the particles \cite[see e.g.][ and references therein]{shalchi09}. 
On larger scales, transient structures like coronal mass ejections (CMEs) or corotating interaction regions (CIRs) perturb the solar wind, making the IMF deviate strongly from a nominal configuration \citep{parker58}.
Since SEPs are tied to the magnetic field lines through the Lorentz force, any alteration of the IMF affects the observed SEP event characteristics.  
In this paper, we focus on the large-scale effects of a solar wind with an embedded CIR, and in particular we study how a non-nominal configuration of the IMF influences the spread of  energetic particles in the heliosphere. 

Understanding the spatial variation of energetic particle  intensities has been the subject of several multi-spacecraft studies \citep[e.g.][to cite a few amongst the more recent works]{dresing12,lario13,dresing14,wiedenbeck13,richardson14,klassen16,lario17}.
These studies illustrated that the spatial dependence of intensity distributions can vary from nominal Gaussian-like shapes  \citep[e.g.][]{richardson14,wiedenbeck13,lario13} to more complex and possibly non-symmetric profiles \citep[e.g.][]{klassen16}. 
 In addition, the longitudinal width of SEP events has been observed to vary significantly across different SEP events, of which some events are circumsolar \citep[e.g.][]{gomez-herrero15}.
Also the onset time of SEP events have been shown to vary significantly across multiple spacecraft, and the nominally best connected spacecraft to the parent particle source do not always show the earliest onset \citep[e.g.][]{klassen15}.  
Various scenarios have been proposed to explain these  observations,  ranging from processes occurring in the corona, such as strongly tilted non-radial magnetic fields \citep[e.g.][]{klein08,klassen18} or  asymmetric coronal shocks waves \citep[e.g.][]{lario14}, to processes occurring in the interplanetary medium such as meandering field lines  and/or different pitch-angle  %strongly varying 
scattering conditions along  particle paths residing in different solar wind streams \citep[e.g.][]{pacheco17}. 
 In addition, as discussed in the previous paragraph, strong deviations of the IMF from the nominal Parker spiral due to transient structures such as CMEs or CIRs may also significantly alter the characteristics of SEP events, and hence contribute to the spatial variations observed across SEP events.

Recently, \cite{wijsen19} introduced a new three-dimensional particle transport model that solves the FTE by assuming a background solar wind generated by the magnetohydrodynamic (MHD) heliospheric model European heliospheric forecasting information asset (EUHFORIA) \citep{pomoell18}. 
Such a coupling between a three-dimensional MHD solar wind model and a particle transport model is necessary to improve our understanding on
%can be used to obtain a better understanding of
the effect of large-scale solar wind perturbations on SEP transport. 
A similar coupling philosophy has been pursued previously; see for example \cite{kocharov09},  \cite{schwadron10}, \cite{kozarev10}, \cite{schwadron14}, and \cite{wei19}.
Since the solar wind often contains plasma streams of varying speed that may evolve into CIRs, understanding energetic particle transport in such conditions has received much attention in recent years \citep[see e.g.][]{giacalone02,03kocharov,mason12,wu14}.
In \cite{wijsen19}, we used EUHFORIA to model a localised fast  wind stream embedded in an ambient slow solar wind, resulting in a self-consistently generated CIR bounded by  forward and reverse shock waves. 
Using a fleet of virtual satellites distributed in the solar equatorial plane, we showed how the time-intensity profiles of an impulsive SEP event can strongly vary from one observer to the other, even when they are located close to each other, that is, separated by less than $10^\circ$ in longitude.
These differences could only be attributed to the varying solar wind conditions, since the particles were uniformly injected from a source region, situated at the inner boundary of EUHFORIA, which is located at a heliocentric radial distance of 0.1 AU.
 Because of the non-nominal IMF topology, nearby located observers can  magnetically be connected to solar wind regions with very different properties and hence sample different particle distributions. 

The strong dependence of the modelled time-intensity profiles on the position of the observer %spatial coordinate,
suggests that the location and size of the particle source region can also have a considerable effect on the SEP distributions in the heliosphere. 
In a non-nominal solar wind, small shifts in longitude or latitude can  connect the particle source region  magnetically to parts of the solar wind with considerably different characteristics.  
Hence,  any shifts in the location of the source region can alter the energy spectrum, duration, and spatial spread of energetic particle events.
 To get better insight on the importance of this effect we model, in this work, the transport of particles injected from four different source regions located also at 0.1 AU from the Sun, that is, at the inner boundary of our solar wind model.%, i.e., at a heliospheric radial distance of 0.1 AU.
 These particle source regions are chosen to be identical in size and shape, but shifted in longitude and hence  located in different solar wind conditions.  
We use the same solar wind structure as in \cite{wijsen19} (hereafter, Paper~I).
Our results illustrate how the location of the particle source region and hence the IMF topology can strongly alter the resulting particle distributions in all three spatial dimensions of the heliosphere. 

The paper is structured as follows. Section~\ref{sec:modelling} briefly discusses the transport equation solved by our model. 
Next, in Section~\ref{sec:results}, we describe the set-up of our simulations and the results. 
We show how the particle intensities and anisotropies along selected magnetic field lines strongly depend on the solar wind properties encountered by the IMF line. 
Moreover, we illustrate how the longitudinal intensity and anisotropy profiles are modulated by the CIR structure, and how the radial evolution of the longitudinal width of the particle intensity distributions reflects the IMF topology in the solar equatorial plane. 
We analyse this using simulations both with and without cross-field diffusion. 
We end the section by illustrating how the particle intensities have a strong latitudinal dependence as well. 
A summary is given in Section~\ref{sec:conclusion}.

\section{Modelling of SEP transport}\label{sec:modelling}
To study energetic particles in the inner heliosphere, we model the evolution of the  gyro-averaged phase-space distribution function, $f(\vec{x},p,\mu,t)$, using the FTE. As detailed in Paper I,  
%Here,
$\vec{x}$ denotes the phase-space spatial coordinate and $t$ the time, both measured in an inertial frame, whereas the cosine of the pitch angle $\mu$ and the momentum magnitude $p$ are expressed in a frame co-moving with the solar wind.
The FTE  can be formulated either as a time forward or time backward Kolmogorov equation. 
The equivalence between both formulations  follows from the solenoidal condition of the phase-space velocity field, i.e. $\left({d\vec{x}}/{dt},{d\vec{p}}/{dt}\right)$ \citep[see e.g.][]{06zhang}. 
In this work, we use the formalism of time forward It\^o   stochastic differential equations to obtain a solution of the FTE, and hence it is natural to look at the time forward Kolmogorov formulation, expressed in terms of the the directional particle intensity
%particle differential flux
$j(\vec{x},p,\mu,t)$. 
More specifically, 
%the particle differential flux 
$j$ is defined as the number density of particles in phase-space element $2 \pi d\vec{x}dpd\mu$, and is therefore related to the gyro-averaged particle distribution function $f(\vec{x},p,\mu,t)$ through $j =  p^2f$. 
The FTE as a time forward Kolmogorov equation can then be formulated as
\begin{equation}\label{eq:fte}
\begin{aligned}
\pd{j}{t} &+\pd{}{\vec{x}}\cdot\left[\left(\td{\vec{x}}{t}+\pd{}{\vec{x}}\cdot \vec{\kappa}_\perp\right)j\right]+\pd{}{\mu}\left[\left(\td{\mu}{t}j+ \pd{D_{\mu\mu}}{\mu}\right) j\right] \\+& \pd{}{p}\left(\td{p}{t}j\right) = \pdd{}{\mu}\left[D_{\mu\mu}j\right] + \pd{}{\vec{x}}\cdot\left[\pd{}{\vec{x}}\cdot\left(\bm{\kappa}_\perp j\right)\right],
\end{aligned}
\end{equation}
with
\begin{eqnarray}
\td{\vec{x}}{t} &=& \vec{V}_{\rm sw}+ \mu \varv \vec{b} +\vec{V}_d  \label{eq:fte_x}  \\
\td{\mu}{t}&=&\frac{1-{\mu}^2}{2}\Bigg(\varv\nabla\cdot\vec{b}+ \mu \nabla\cdot\vec{V}_{\rm sw}- 3 {\mu} \vec{b}\vec{b}:\nabla\vec{V}_{\rm sw}  \label{eq:fte_mu}\\ 
\notag %added to not have the number just for a term in the equation
&&- \frac{2}{{\varv}}\vec{b}\cdot\td{\vec{V}_{\rm sw}}{t}\Bigg) \\
\td{p}{t} &=& \Bigg(\frac{1-3{\mu}^2}{2}(\vec{b}\vec{b}:\nabla\vec{V}_{\rm sw}) - \frac{1-{\mu}^2}{2}\nabla\cdot\vec{V}_{\rm sw} \label{eq:fte_p} \\
\notag %added to not have the number just for a term in the equation
&&-\frac{{\mu} }{{\varv}}\vec{b}\cdot\td{\vec{V}_{\rm sw}}{t}\Bigg){p}.
\end{eqnarray}
In this equation 
$\varv$  is the particle speed,
$\vec{V}_{sw}$ is the solar wind velocity, and
$\vec{b}$ the unit vector in the direction of the mean magnetic field,
$D_{\mu\mu}$ is the pitch-angle diffusion coefficient, 
$\bm{\kappa}_\perp$ the spatial cross-field diffusion tensor, and
$\vec{V}_d$ the drift velocity due to the gradient and curvature of the mean magnetic field. 
The choice of the diffusion terms was explained in detail in Paper~I.  
For referential convenience, we restate the definition of the cross-field diffusion tensor as follows: 
\begin{eqnarray}\label{eq:cross_field}
\bm{\kappa}_\perp &=&\frac{\pi }{12}{\alpha\varv \lambda_\parallel}\frac{B_0}{B}\left(\mathbb{I}-\vec{bb}\right),
\end{eqnarray}
where $\lambda_\parallel$ is the particle mean free path,  $\mathbb{I}$ is the unit tensor, $\vec{bb}$ is a dyadic product of the magnetic field unit vectors, and $\alpha$ is a free parameter that determines the ratio of the parallel to perpendicular mean free path at a reference magnetic field strength $B_0$. 
The details of our numerical procedure for solving the transport model are described in Paper~I. In the next sections, we show the particle differential intensity defined as $ I = \frac{1}{2} \int_{-1}^{1} j(\vec{x},E,\mu,t)\,d\mu$, and the parallel first order anisotropy as $A = 3 \int_{-1}^{1} \mu j(\vec{x},E,\mu,t)\,d\mu / \int_{-1}^{1} j(\vec{x},E,\mu,t)\,d\mu$. 
%\cite{wijsen19}.

\section{Solar energetic particle transport near a CIR}\label{sec:results}
\begin{figure*}
    \centering
    \includegraphics[scale=0.27]{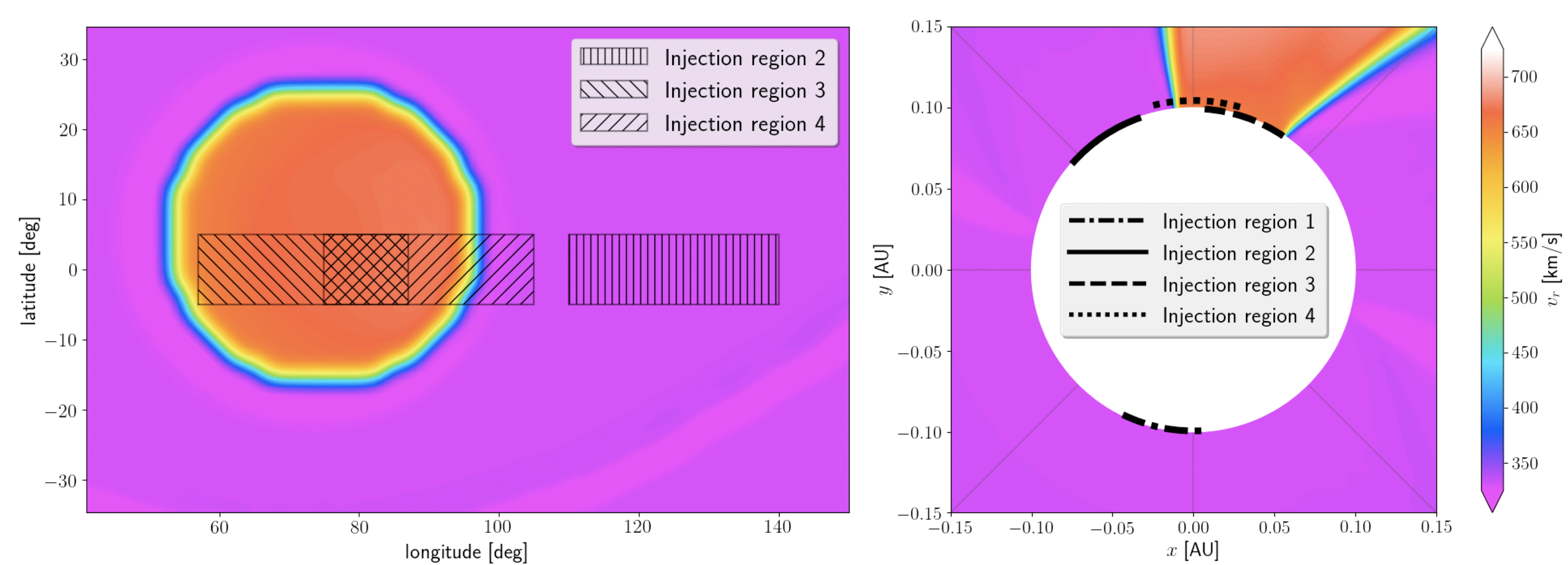}
    \caption{Snapshot of the solar wind radial velocity showing a part of the $r=0.1$~AU inner boundary (left panel) and a part of the solar equatorial plane (right panel). Indicated on both panels are the different particle injection regions.} 
    \label{fig:injection_regions}
\end{figure*}
In Paper I, we used EUHFORIA to model a solar wind with an embedded fast solar wind stream, generating a CIR bounded by a forward and reverse shock wave. 
In this work, we use the same solar wind model, yet we inject protons uniformly in the four different injection regions illustrated in Fig.~ \ref{fig:injection_regions}.   
These four injection regions are all centred around the solar equatorial plane and have a longitudinal width of $30^\circ$ and a latitudinal width of $10^\circ$. 
This rectangular shape makes it straightforward to quantify the latitudinal and longitudinal deformation of the particle streaming zone  due to a non-nominal IMF, since in a Parker spiral the longitudinal and latitudinal width remain constant and hence, the rectangular shape is exactly preserved.    
For the first case, the injection region is located in the slow solar wind (see the right panel of Fig.~\ref{fig:injection_regions}) far enough from the CIR such that the particles travel in a nominal slow solar wind. 
For case~2, the injection region is located in the slow solar wind in front of the CIR in such a way that magnetic field lines connect this injection region with the forward shock of the CIR. 
In contrast, injection region~3 is entirely located in the fast solar wind stream and magnetically connected to the reverse shock of the CIR. 
%such that the magnetic field lines connect this  injection region  with the reverse shock of the CIR. 
Finally, the injection region of case~4 is the same injection region as discussed in Paper~I, and is located partly in the slow solar wind and partly in the fast solar wind. 
Like in Paper~I, we inject protons impulsively (i.e. a delta injection in time) and with an initial energy of 4 MeV to facilitate the tracking %keeping track 
of the energy changes of the particles in the solar wind. 
We assume the same scattering conditions as in Paper~I, which are summarised as follows.
 The pitch-angle diffusion coefficient is characterised similarly to \cite{Agueda08}, and we take the proton radial mean free path constant and equal to $\lambda_r^\parallel = 0.3$ AU for 4 MeV protons.  
The values of the parameters characterising the cross-field diffusion coefficient (Eq~\ref{eq:cross_field}) are $\alpha=10^{-4}$ and {$B_0 = \max B(r=1\rm \, AU) = 9.7 \, \rm nT$}. 
As explained in Paper~I, this maximum value of the magnetic field is obtained in the compressed shocked slow solar wind, where the cross-field diffusion is thus minimal.
%and hence the cross-field diffusion is minimal there. 
In our simulations, the ratio of the perpendicular mean free path to the parallel mean free path inside the CIR and at 1 AU is  at most $\lambda_\perp/\lambda_\parallel = 3.81\times10^{-4}$.  
This is small compared to the values obtained by \cite{dwyer97}, who find
ratios of the order of unity for three CIRs using 44\,--\,313~keV/nuc helium measurements from the
Wind spacecraft. 
Therefore, we expect that the effects of the cross-field diffusion near the CIR found in the results below are likely to be enhanced with a more realistic treatment of the cross-field motions.

Finally, we would like to note that 
the finite resolution of our MHD simulation produces  shock waves that are wider than real interplanetary CIR shocks.  
Therefore, as explained in paper~I, we can argue that 
%making use of 
the terminology ``particle shock acceleration'' is not strictly applicable for particle acceleration near the strong compression waves in our simulations.   
However, if the particle mean free path across the shock is significantly larger than the width of the  high-amplitude compression waves (as happens in our simulations), then these waves act on the particles as shocks. 
 More specifically, the particles gain energy due to their motion and scattering in  converging flows characterising such compression waves,  which is analogous to what happens during first-order Fermi shock acceleration \cite[see also ][for a discussion]{giacalone02}.

\subsection{Particle intensities along a magnetic field line}
\begin{figure*}
    \centering
    \includegraphics[scale=0.49]{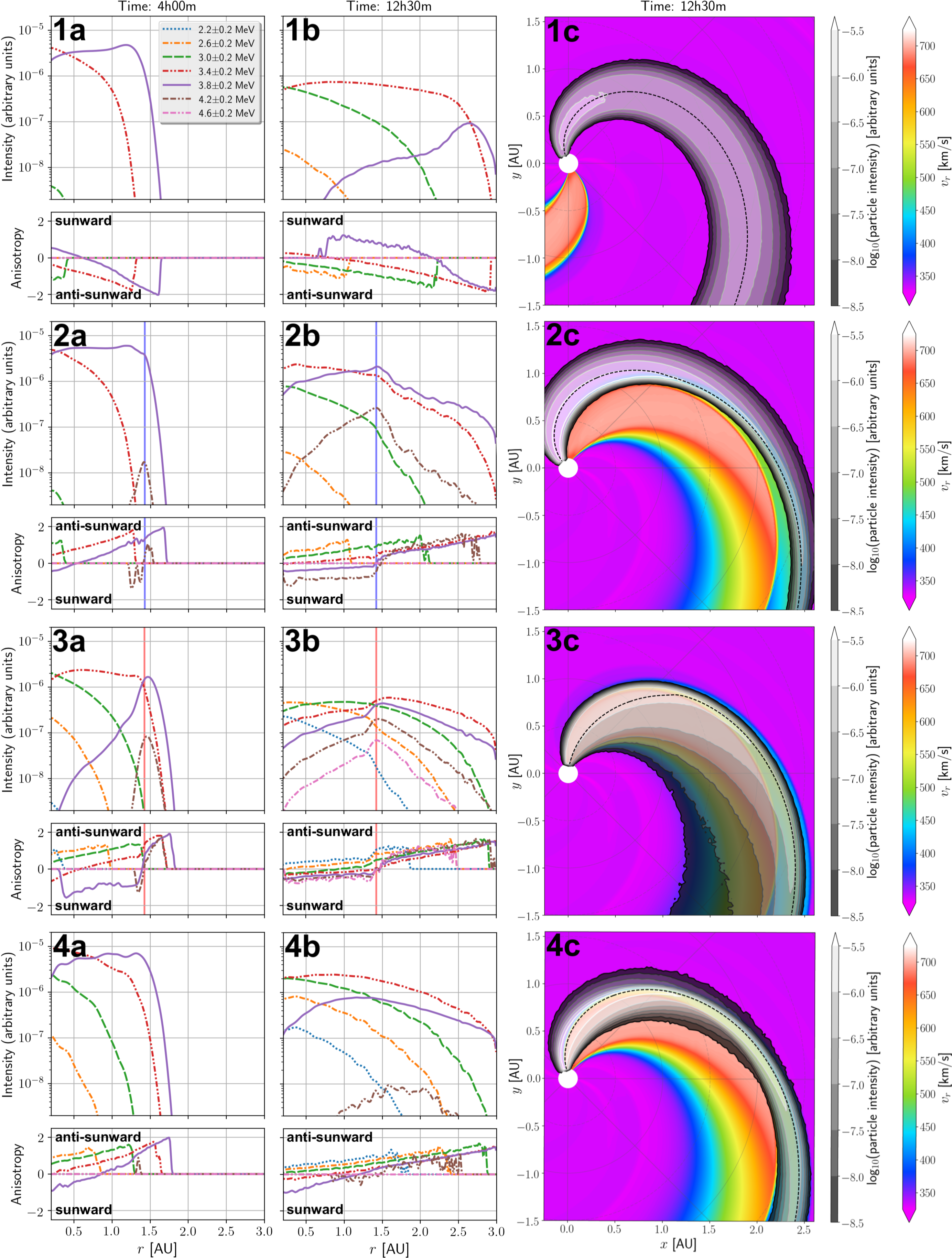}
    \caption{Columns~a and ~b show snapshots of the particle intensities and anisotropies along a selected magnetic field line (see text). This IMF line is indicated by a dashed line in the figures of the column~c. Also shown in column~c are the integrated particle intensities drawn in grey shades on top of the radial velocity profile of the solar wind. Each row shows the results for each injection region labelled by their number %corresponds to a different injection region, with the row number equal to the number of the injection region 
    (see  Fig.~\ref{fig:injection_regions}). The blue and red vertical lines indicate the forward and reverse shocks, respectively. } 
    \label{fig:field_lines}
\end{figure*}

We now analyse the particle intensities and anisotropies along four magnetic field lines. 
More precisely, for each injection case we select a different magnetic field line to illustrate how the varying solar wind conditions may produce particle intensity and anisotropy profiles that are substantially different from one field line to the other. 
The results are shown Fig.~\ref{fig:field_lines} for the simulations with non-zero cross-field diffusion. Columns~a and~b of the figure show the intensity and anisotropy profiles along the IMF line as a function of the radial coordinate 4 hours and  12.5 hours, respectively, after the particle injection.
In the panels showing the anisotropy profiles, we include the particle propagation direction since the magnetic field polarity  in the unperturbed solar wind and near the fast solar wind stream is different (see also Paper~I). 
For injection region~1, we choose an IMF line centred in the particle streaming zone, as illustrated in panel~1c of Fig.~\ref{fig:field_lines}. 
Panels~1a and~1b show that, as a result of adiabatic deceleration, the lower energy channels are quickly populated by the particles. For example, after $\sim 4$ hours, the $3.4\pm0.2$ MeV is already showing intensities of the same order of magnitude as the $3.8\pm0.2$ MeV energy channel.
This is especially true close to the Sun, where for a Parker spiral, the adiabatic deceleration is the strongest since $\nabla\cdot \vec{V}_{sw} \sim V_{sw}/r$. 
We note that, owing to our mono-energetic injection of 4 MeV particles and the nominal solar wind conditions, the $3.8\pm0.2$ MeV energy channel can only lose particles, since it cannot get replenished through adiabatic deceleration of protons with energies above 4 MeV or through proton acceleration in converging flows for instance.
The $3.8\pm0.2$ MeV  protons remaining a prolonged time close to the Sun move quickly to a lower energy channel, since adiabatic deceleration is most efficient at small radial distances.
Hence, protons can only remain in the $3.8\pm0.2$ MeV if they propagate quickly, and hence with a pitch angle close to 180$^\circ$, to larger radial distances where the adiabatic deceleration is weaker. This, combined with the focusing effect, explains the high anti-sunward anisotropies of the $3.8\pm0.2$ MeV energy channel at the start of the event. 
At a later time, when the bulk of $3.8\pm0.2$ MeV protons has reached farther radial distances, the intensity of this particles drops and the anisotropies switch signs, as  is shown in panel~1b around $r\sim2.25$ AU. 
 For $r<2.25$ AU,  the measured sunward-streaming protons only reversed their direction through scattering after having travelled to large radial distances where the adiabatic deceleration is weaker.
However, we note that if they keep travelling in the sunward direction long enough, they eventually populate a lower energy channel.  
The combined high efficiency of adiabatic deceleration and focusing at small radial distances also explains why the lower energy channels typically show strong anti-sunward anisotropies, even 12.5 hours after the delta injection (see e.g. the $3.0\pm0.2$ MeV  channel the panel~1b). 
This is because close to the Sun, adiabatic deceleration continuously and efficiently decelerates particles to these lower energy channels and focusing increases $|\mu |$; thus, resembling the effect of a time-extended particle injection in those channels.
%, reminiscent to an extended particle injection in those channels. 

In addition, our simulations also show relatively strong anisotropies  because we use a rather fine energy resolution $\Delta E = 0.4 $ MeV. 
If we were to consider, for example a $3.0\pm 1.0$ MeV energy channel instead,  then the anisotropy in panels~1b would be relatively small for $r<1.5$ AU, since it would be modulated  by the energy channel with highest intensity, i.e. the $3.4\pm0.2$ MeV  channel in this case. 
This however means we would lose all information about the behaviour of particles with energies outside the $3.4\pm0.2$ MeV channel. Finally we note that high anisotropies obtained with corresponding low intensities are not very significant when comparing with in situ observations, since these anisotropies would likely be mitigated by an isotropic background of energetic particles.

Next, we look at a magnetic field line originating from the centre of injection region~2, as illustrated in the panel~2c of Fig.~\ref{fig:field_lines}. 
This IMF line crosses the forward shock wave of the CIR at a heliospheric radial distance of $r\sim 1.4$ AU. 
This is reflected in the intensity profiles of the snapshot after 4 hours (panel~2a), through the appearance of particles in the energy channel $4.2\pm0.2$ MeV, centred on the forward shock (indicated by the blue vertical line in the panels 2a and 2b of Fig.~\ref{fig:field_lines}). 
Apart from this, the intensities at this time  are not very different as compared to the case of injection in the unperturbed solar wind, as expected. 
This changes when examining the intensity profiles after 12.5 hours (panel~2b), since more particles have had time to interact with the forward shock and to travel inside the CIR. 
The $4.2\pm0.2$ MeV energy channel, peaking at the forward shock, is now substantially populated.
We note that the $3.8\pm0.2$ MeV energy channel,  which in nominal conditions (case~1) only loses particles, now gets replenished through particle acceleration near the shock. 
In addition,  the $3.8\pm0.2$ MeV  energy channel can  receive particles  away from the shock, through adiabatic deceleration of particles that have been accelerated into the $4.2\pm0.2$ MeV energy channel at the forward shock. 
Together, this explains why  the $3.8\pm0.2$ MeV energy channel shows high intensities along the entire field line, in sharp contrast to the case of the unperturbed solar wind. 
Finally, we note that the anisotropy profiles in panel~2b have also been strongly altered in comparison to the nominal case shown in panel~1b. 
Especially the sign-switch of the anisotropies of energy channels  $3.8\pm0.2$ MeV  and $4.2\pm0.2$ MeV at the forward shock is notable, and it is a consequence of the accelerated particles streaming away from the shock in both sunward and anti-sunward direction.

The third row of Fig.~\ref{fig:field_lines} shows the particle intensity and anisotropy along a field line originating from the centre of injection region 3. 
As in case~2, the field line enters the CIR at a heliospheric radial distance of $r \sim 1.4$ AU. 
However, for this case, the IMF line originates in the fast solar wind and crosses the reverse instead of the forward shock upon entering the CIR.
The snapshot after 4 hours (panel~3a) is considerably different from the two previous snapshots, since the  IMF line resides in the fast solar wind for $r<1.4 $ AU, instead of the slow solar wind. 
As a consequence,  the particles experience considerably more adiabatic deceleration since the divergence of the solar wind velocity is larger in the fast than in the slow solar wind (see Paper~I for a discussion). 
This explains why energy channel $3.8\pm0.2$ MeV is quickly depleted of particles close to the Sun, whereas the lower energy channels show much higher intensities compared to the previous two cases. 
We also note that energy channel $4.2\pm0.2$ MeV starts to get populated with particles crossing the reverse shock wave (see the brown profile peaking close to the red vertical line). 
Also the snapshot after 12.5 hours (panel~3b) is  strongly different as compared to the previous two cases, reflecting the interplay between strong adiabatic deceleration at small radial distances ($r<1.4 $ AU) and acceleration at the reverse shock. 
Comparing with the snapshot after 4 hours illustrates that the reverse shock has repopulated the $3.8\pm0.2$ MeV energy at small radial distances. 
In addition, acceleration at the reverse shock has filled both the $4.2\pm0.2$ MeV and the $4.6\pm0.2$ MeV energy channels. Comparing with  case 2 (panel~2b), we remark that the $4.6\pm0.2$ MeV energy channel was not populated for the IMF line  crossing the forward shock at the same radial distance, indicating that the reverse shock is a more efficient accelerator in our simulation. 
Apart from this, we find again that the anisotropies of the highest energy channels, this time including the $3.4\pm0.2$ MeV channel, switch sign near the shock, indicating that particles are accelerated there.

Finally, we look at a magnetic field line originating close to the centre of injection region~4, as illustrated in the last panel of the fourth row of Fig.~\ref{fig:field_lines}. 
At a small heliospheric radial distance (${\sim} 0.6$ AU), this IMF line enters the CIR, crossing the  compression wave that steepens into the reverse shock wave at larger radial distances.
The compression wave is not yet strong enough to accelerate particles into the $4.2\pm0.2$ MeV energy channel, yet it contributes by counteracting to some extent the adiabatic deceleration that the particles undergo while travelling in the fast solar wind. 
This can be seen by noting that the $3.8\pm0.2$ MeV energy channel remains substantially populated, similar to cases 2 and 3 (e.g. panels~2b and~3b) and in contrast to case 1 (e.g. panel~1b). On the other hand, particles residing for some time in the fast solar wind, away from the compression wave, are considerably decelerated. 
As a result, the $3.0\pm0.2$ MeV and the $2.6\pm0.2$ MeV energy channels show  intensities similar to the previous case, which is in contrast to the first two cases.
Panel 4b of Fig.~\ref{fig:field_lines} % The intensities  at t=12h30
shows the presence of a small population of accelerated particles in the $4.2\pm0.2$ MeV channel. However, these particles are not seen along this IMF line for the simulation done with zero cross-field diffusion (not shown here). Therefore, these accelerated particles likely originate from adjacent field lines that cross the reverse shock wave at ${\sim}$1.4~AU, as indicated by the positive anisotropies observed at larger distances.

%\textcolor{violet}{You have done a very good summary of the main characteristics seen in these plots. I liked it. My comments are more related to the next work on particle acceleration. Here we are not indeed studying the acceleration mechanisms, but point to that particles are accelerated close to or at the shock waves.}

\subsection{Longitudinal intensity profiles}

\begin {table}
\centering
  \caption{  Longitudinal widths of the particle intensities and the longitudinal increases due to cross-field diffusion.}\label{tab:dlon}
  \begin{tabular}{ c c  c  c  c  c  }
   \hline \hline
    Case   & $\Delta\varphi_i^{0}$  & $\Delta\varphi_i^{\perp}$   & $\Delta\varphi_{\rm{L},i}$    & $\Delta\varphi_{\rm{U},i}$   & $\Delta\varphi_i^{\perp}$ / $\Delta\varphi_i^{0}$   \\ \hline
    1 & $30^\circ$& $66^\circ$ & $17.9^\circ$ & $18.1^\circ$ & 2.2 \\
    2 & $11.1^\circ$ & $36.1^\circ$  &$5.1^\circ$  & $19.9^\circ$ & 3.2 \\ 
    3 & $47.3^\circ$ & $75.8^\circ$  &$23.8^\circ$ &$4.7^\circ$ & 1.6  \\ 
    4 &  $5.6^\circ$ & $41.2^\circ$  &$15.4^\circ$  & $20.1^\circ$ & 7.3 \\ \hline
  \end{tabular}
 \tablefoot{Numbers are for a heliocentric radial distance $r=1.5$ AU in the solar equatorial plane. }
\end {table}

\begin{figure*}
        \centering
        \begin{tabular}{cc}
        \includegraphics[width=0.4\textwidth]{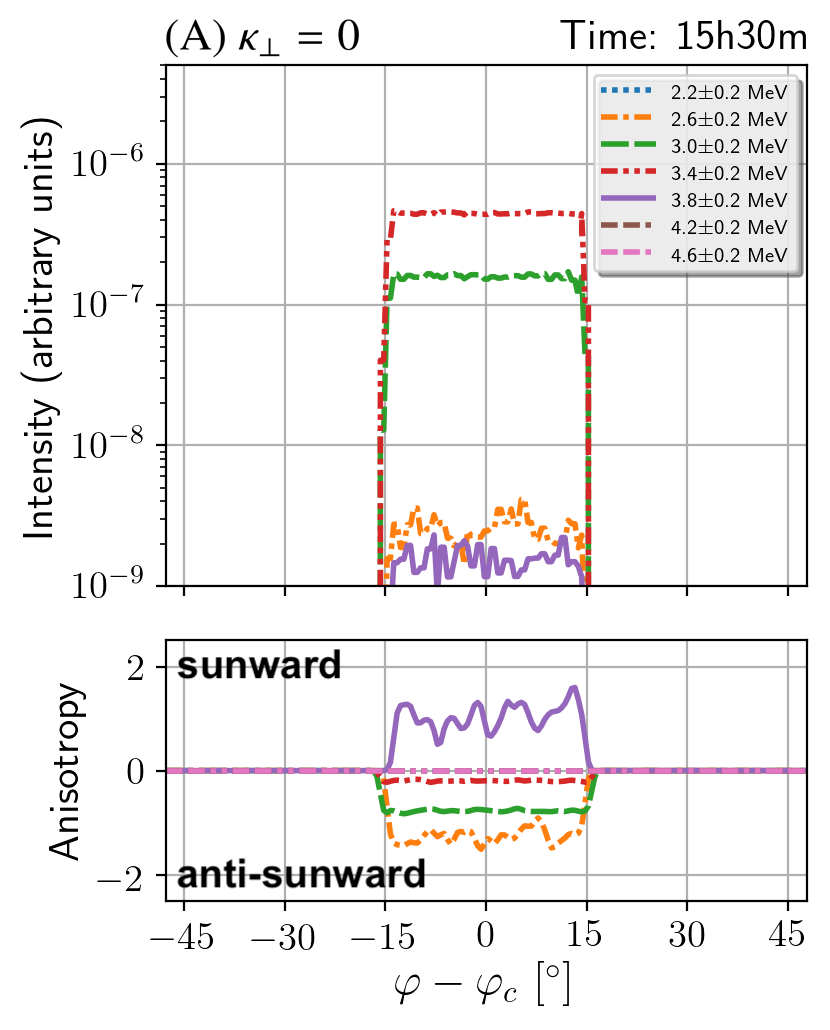}&
        \includegraphics[width=0.4\textwidth]{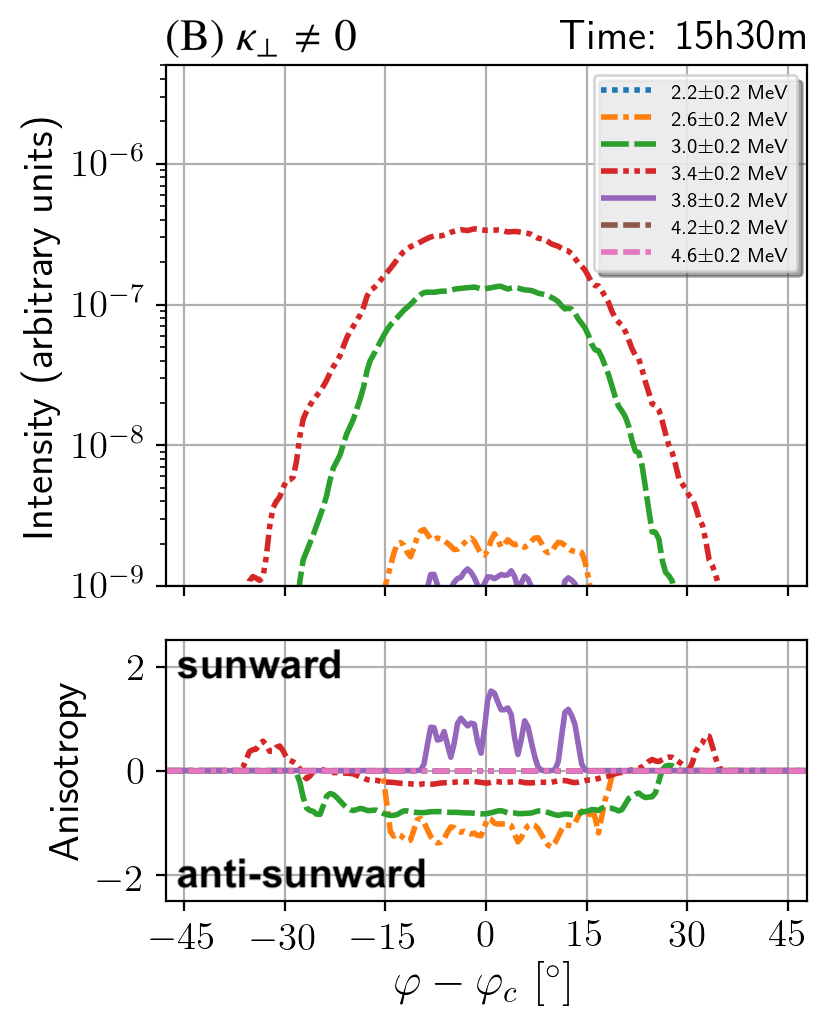}\\
    \end{tabular}
    \caption{Snapshot of the intensities (upper panels) and anisotropies (lower panels) of protons originating from injection region 1, at time 15h30,  both for the case without (left panel) and with (right panel) cross-field diffusion. The profiles are measured in the solar equatorial plane at a heliocentric radial distance of 1.5 AU, and as a function of longitudinal separation from the centre of the injection region.}
     \label{fig:lon_1}
\end{figure*}
In this section, we examine  the variation of the particle intensity and anisotropy profiles in the solar equatorial plane at a fixed heliographic radial distance of 1.5 AU.  For the four different cases, we study the intensities and anisotropies as a function of the longitude separation  from the centre of the injection region, $\varphi - \varphi_c$. Here, $\varphi$ denotes the counterclockwise measured longitude coordinate and $\varphi_c$ represents the longitude of the point at $1.5$ AU that connects magnetically to the centre of the injection region at 0.1 AU. 
Moreover,  the figures discussed below are obtained 15.5 hours after the particle injection, and both the results for  simulations with (panels A) and without (panels B) cross-field diffusion are shown. In order to quantify the longitudinal width of the particle intensity distributions at 1.5 AU for the different cases, we define $\Delta \varphi_i^{0,\perp}$ for case $i$ as the longitudinal width of the region  where the omnidirectional intensity, integrated over all energies, is larger than $10^{-9}$ in our simulations. 
The superscripts $0$ and $\perp$ indicate whether we are considering a simulation with or without cross-field diffusion, respectively.  
If we denote the longitudes of the lower and upper longitudinal  edges  of the particle intensity distribution by $\varphi_{\rm{L},i}$ and $\varphi_{\rm{U},i}$,  respectively, then we can write
$\Delta \varphi_i^{0,\perp} = \varphi_{\rm{U},i}^{0,\perp} - \varphi_{\rm{L},i}^{0,\perp}$. 
The edge-longitudes $\varphi_{\rm{L},i}$ and $\varphi_{\rm{U},i}$ allow us to measure the longitudinal width increase due to cross-field diffusion of the intensity distribution
%the particle streaming zone
as $\Delta\varphi_{\rm{L},i} = \varphi_{\rm{L},i}^{0} - \varphi_{\rm{L},i}^{\perp}$
 and  
 $\Delta\varphi_{\rm{U},i} = \varphi_{\rm{U},i}^{\perp} - \varphi_{\rm{U},i}^{0}$. 
The longitudinal widths $\Delta \varphi_i^{0,\perp}$ of the particle intensities and the longitudinal width increases $\Delta \varphi_{\rm{L,U},i}$ for the four different cases are summarised in  Table~\ref{tab:dlon} and discussed below together with the corresponding figures.  

Figure~\ref{fig:lon_1} shows the longitudinal profiles for injection region~1, for which particles propagate in a nominal solar wind. 
As a consequence, the longitudinal intensity profiles are symmetric around the centre of the particle streaming zone, and the sharp intensity cut-offs at the edges for the case without cross-field diffusion (reflecting the sharp transition in the spatial injection profile) are smoothed for the case with cross-field diffusion. We note that the cross-field diffusion spreads the $3.4\pm0.2$ MeV protons more in longitude than than for example the $2.6\pm0.2$ MeV protons. 
This is partly because the cross-field diffusion increases with the speed of the particles
\footnote{\label{fn:vdependence} As detailed in Paper~I, we have that $\lambda_\parallel^r \propto \varv^{2-q}$, where $q = 5/3$ denotes the exponent of the power spectrum of the magnetic turbulence. Equation~\eqref{eq:cross_field} then gives that $\kappa_\perp \propto \varv^{3-q} $.}, 
but also because the  $3.4\pm0.2$ MeV channel gets populated before the $2.6\pm0.2$ MeV energy channel, and hence the particles have more time to diffuse across the magnetic field. 
From Table~\ref{tab:dlon}, we note that $\Delta\varphi^{0}= 30^\circ$, which means that the longitudinal width of the particle intensity distribution remains equal to the width of the injection zone, as expected for a nominal IMF configuration. Finally, we point out that 
the wiggles in energy channels $3.8\pm0.2$ MeV and $2.6\pm0.2$ MeV are due to the low statistics at those low intensities. 
\begin{figure*}
        \centering
        \begin{tabular}{cc}
        \includegraphics[width=0.4\textwidth]{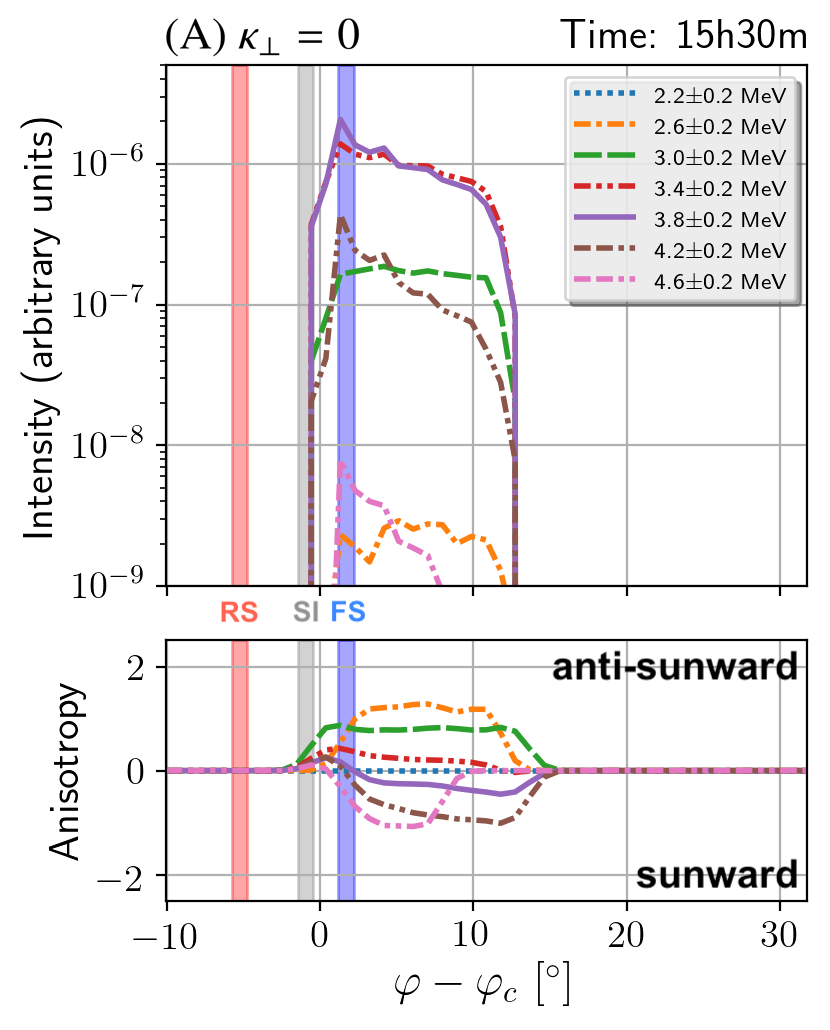}&
        \includegraphics[width=0.4\textwidth]{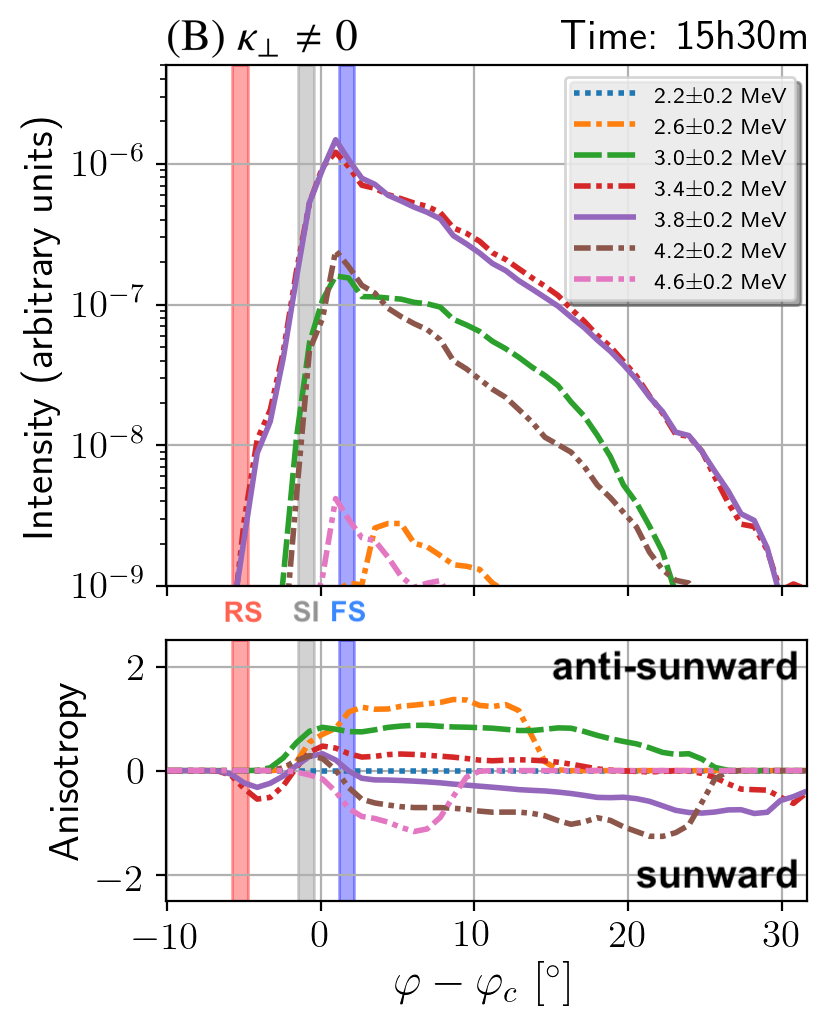}\\
    \end{tabular}
    \caption{Same as Fig.~\ref{fig:lon_1}, but for protons originating from injection region 2. The grey, red, and blue bands are centred on the SI, the reverse shock (RS), and the forward shock (FS), respectively.}
     \label{fig:lon_2}
\end{figure*}

In contrast to case~1, Fig.~\ref{fig:lon_2} shows that for case~2 the intensity profiles are strongly asymmetric around the centre of the particle streaming zone. 
For $\varphi - \varphi_c \lesssim 2^\circ$, particles are travelling in the highly compressed region downstream of the forward shock. Moreover, we note that the intensities peak around the forward shock, where particles get accelerated. 
Because of this shock acceleration the $3.8\pm0.2$ MeV energy channel shows high intensities, which is in sharp contrast to case~1.
Table~\ref{tab:dlon} shows that the  width of the particle streaming zone $\Delta\varphi_2^{0}$ is less than half the width of the previous case, and hence also less than half the longitudinal width of the injection region. 
This decrease in width is due to particles following IMF lines that cross the forward shock, where the magnetic field is compressed; thus, this also explains why the peak intensities in Fig.~\ref{fig:lon_2} are higher than those in Fig.~\ref{fig:lon_1}. 
When looking at panel~B of Fig.~\ref{fig:lon_2}, it is noticeable  that, in contrast to case~1, the effect of cross-field diffusion on  the intensity profiles is asymmetric. 
This can also be seen from Table~\ref{tab:dlon}, which shows that $\Delta\varphi_{\rm L,2} \ll  \Delta\varphi_{\rm U,2} $.
The reason for this asymmetry is that, near $\varphi - \varphi_c \sim 0$, the particles encounter the stream interface (SI), which is characterised by a strong magnetic field, and hence a weaker cross-field diffusion according to Eq.~\eqref{eq:cross_field}. 
Finally, we remark that the cross-field diffusion has a  larger effect on the longitudinal spread for case~2 than for case~1, when comparing  $\Delta\varphi_1^{\perp}$ / $\Delta\varphi_1^{0}= 2.2$ with $\Delta\varphi_2^{\perp}$ / $\Delta\varphi_2^{0} = 3.2$ (see Table~\ref{tab:dlon}).  
Despite the increased magnetic field in the shocked slow solar wind, the  cross-field diffusion can thus still be important since  magnetic field lines that are significantly separated in the unperturbed solar wind  are closely compressed in the CIR. 
Hence a small cross-field diffusion in the CIR is sufficient to transport the particles across these IMF lines, and if afterwards the particles return to the unperturbed solar wind, they produce a wide particle streaming region. 
\begin{figure*}
        \centering
        \begin{tabular}{cc}
        \includegraphics[width=0.4\textwidth]{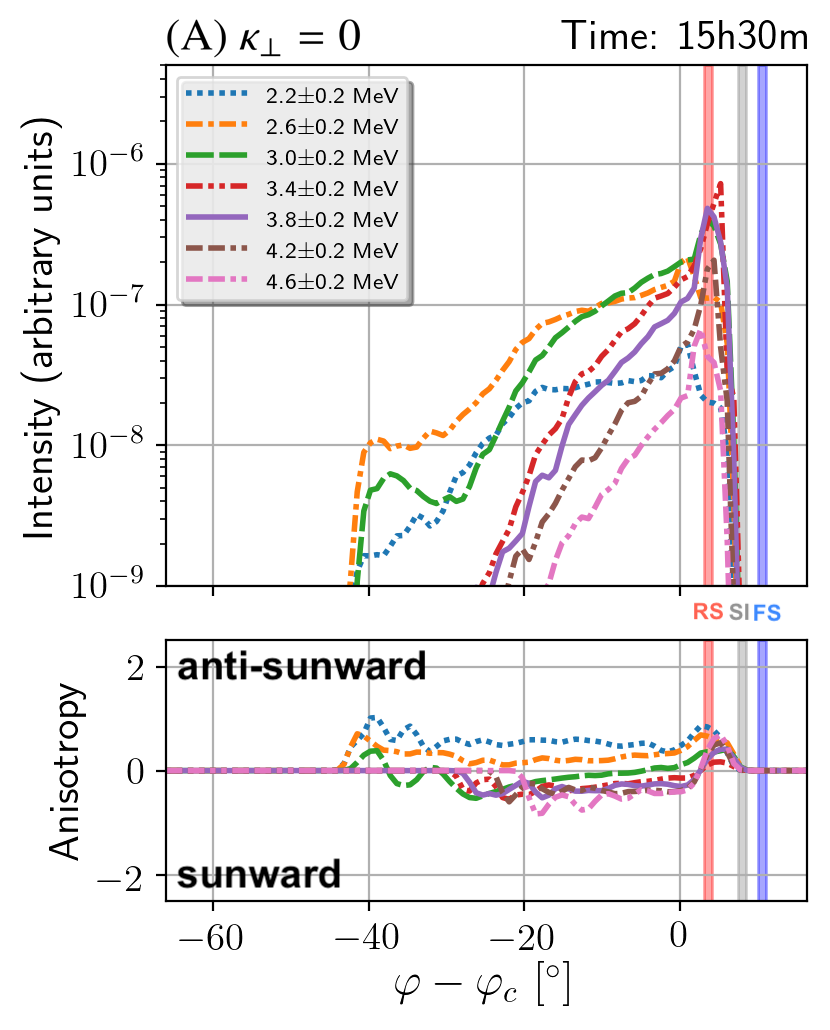}&
        \includegraphics[width=0.4\textwidth]{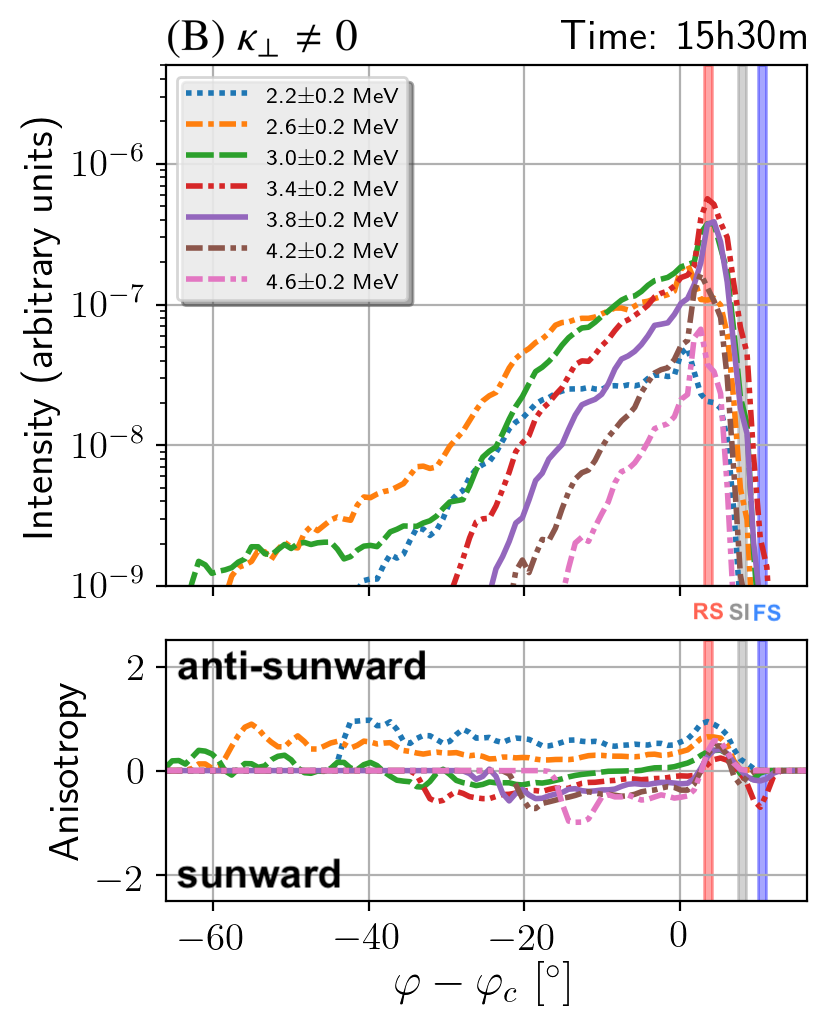}\\
    \end{tabular}
    \caption{Same as Figs.~\ref{fig:lon_1} and \ref{fig:lon_2}, but for protons originating from injection region 3.}
     \label{fig:lon_3}
\end{figure*}

The longitudinal intensity and anisotropy profiles for case~3 are illustrated in Fig.~\ref{fig:lon_3}. 
We see that the intensities for this case peak around the reverse shock, which also shows the presence of  accelerated particles populating the energy channels $4.2\pm0.2$ MeV and $4.6\pm0.2$ MeV. 
Similar to case~2 the intensity profiles are strongly asymmetric around the centre of the particle streaming zone. 
This is because for $\varphi - \varphi_c \gtrsim 5^\circ$ particles are travelling in the highly compressed region downstream of the reverse shock, whereas for $\varphi - \varphi_c \lesssim -30^\circ$, particles are travelling in the rarefaction region behind the fast solar wind stream. 
From Table~\ref{tab:dlon}, we note that $\Delta\varphi_3^0 > 30^\circ$, that is, even without cross-field diffusion, the longitudinal width of the particle intensity distribution at 1.5 AU is larger than the original width of the injection region.
This can be  attributed to the rarefaction region behind the fast solar wind stream.  In addition, we note that $\Delta\varphi_3^0 $ would have even been much larger if it were not for the compressed IMF in the CIR at the upper longitude edge of case~3.
%Moreover, note also the local minimum of the $3.0\pm0.2$ MeV channel around $\varphi - \varphi_c  \sim -30^\circ$. The lower solar wind speed in the rarefaction region as compared to the fast solar wind stream, implies that particles undergo less adiabatic deceleration, which in turn explains the larger intensity found in the  $3.0\pm0.2$ MeV channel at $\varphi - \varphi_c \lesssim -30^\circ$.
%This is because the lower solar wind speed in the rarefaction region  implies a lower adiabatic deceleration, explaining the larger intensity in the  $3.0\pm0.2$ MeV channel. 

Looking at panel~B of Fig.~\ref{fig:lon_3}, we see that similar to case~2, the cross-field diffusion has a clear asymmetric effect on the intensity profiles. 
This asymmetry is also evident from Table~\ref{tab:dlon}, showing $\Delta\varphi_{\rm L,3} \gg \Delta\varphi_{\rm U,3} $, and can be attributed to the proximity of the SI to the upper boundary of the particle longitudinal spread. 
In the CIR, only few particles diffuse through the SI and hence the SI is characterised by a steep drop in intensities.
%\textbf{Despite the compressed IMF in the CIR and the weak cross-field diffusion near the SI, we note from Table~\ref{tab:dlon} that $\Delta_3^\perp = 75.8$, meaning that the streaming zone has increased by a factor of $\sim 2.5$ when comparing with the width of the injection region. } 
We remark that $\Delta\varphi_3^{\perp}$ / $\Delta\varphi_3^{0} \sim 1.6 $ (see Table~\ref{tab:dlon}) is smaller than for the previous two cases, despite the fact that the cross-field diffusion is the strongest in the fast stream and the rarefaction region owing to the weak IMF in these regions. The latter is confirmed in Table~\ref{tab:dlon} by noting that $\Delta\varphi_{L,3}$ is the largest of all  $\Delta\varphi_{L,i}$ and $\Delta\varphi_{U,i}$. 
%However, in those regions, the particles need to transverse a significantly larger cross-field distance to end up on a field-line that has a large footpoint separation from its original field line. In contrast, we might expect that the effect of cross-field diffusion will be much more notable at large radial distances due to the strongly diverging character of the IMF in the rarefaction region.
\begin{figure*}
        \centering
        \begin{tabular}{cc}
        \includegraphics[width=0.4\textwidth]{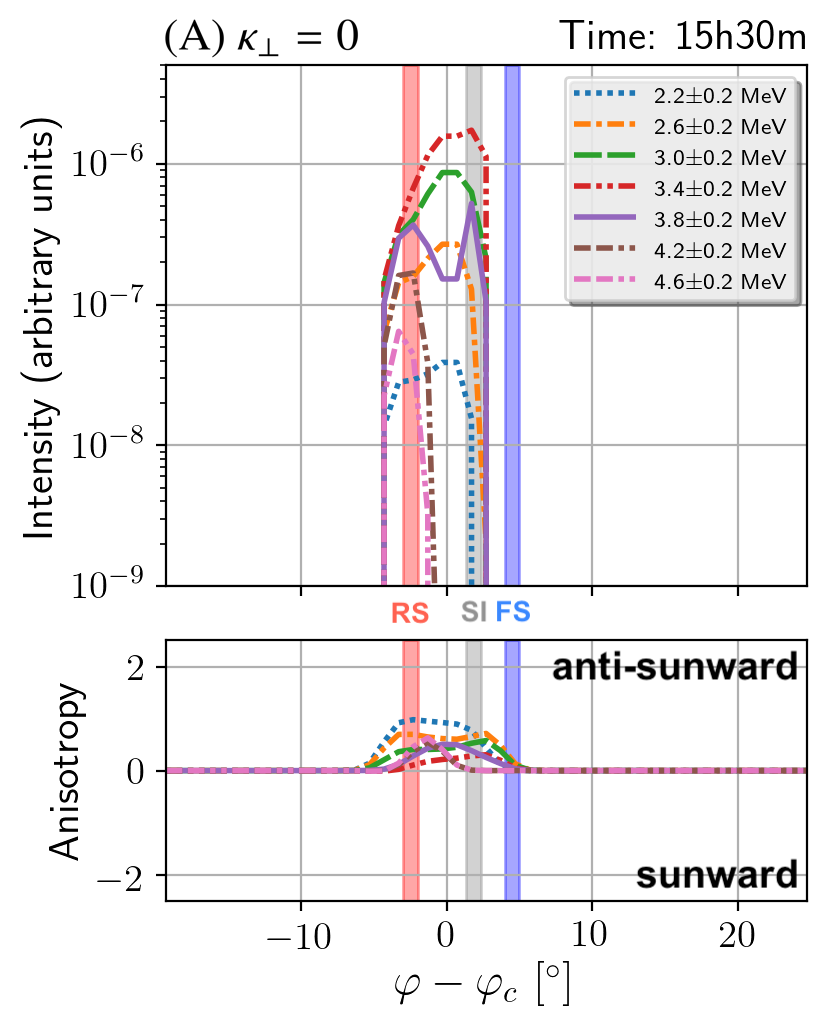}&
        \includegraphics[width=0.4\textwidth]{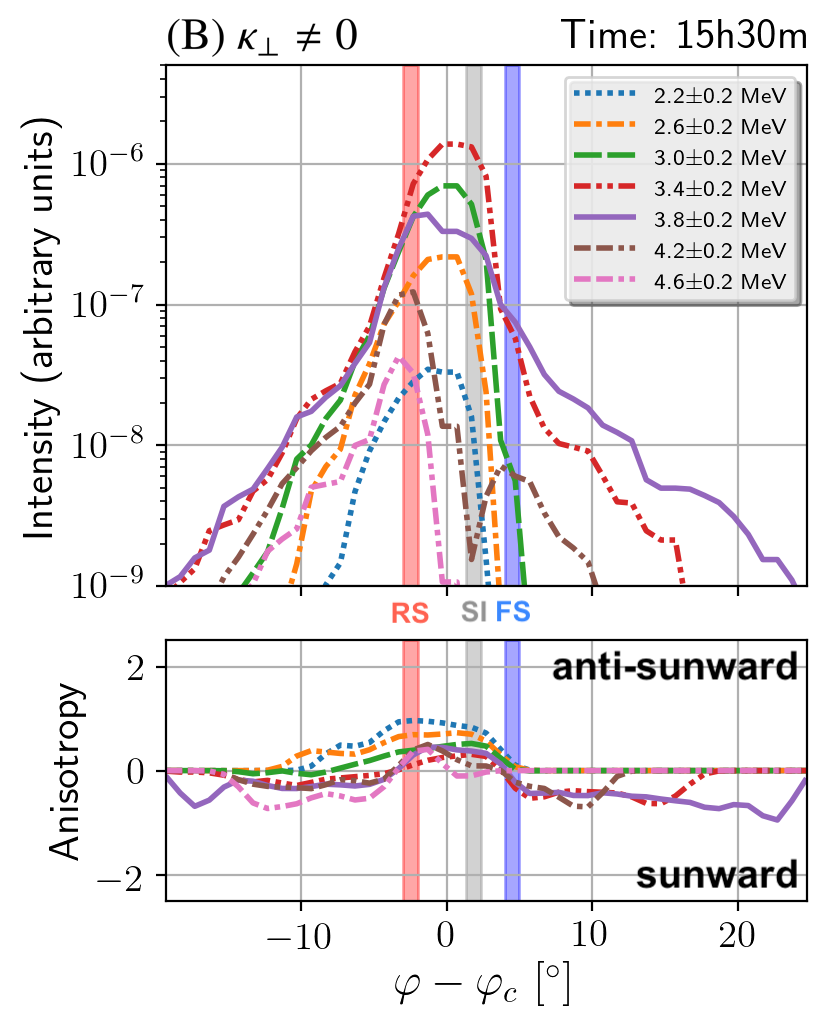}\\
    \end{tabular}
    \caption{Same as Figs.~\ref{fig:lon_1} and \ref{fig:lon_2}, but for protons originating from injection region 4.}
     \label{fig:lon_4}
\end{figure*}

The longitudinal particle intensity and anisotropy profiles for case~4 are shown in Fig.~\ref{fig:lon_4}.
Panel~A of this figure illustrates that  the streaming zone is much narrower compared to the previous cases, reflecting the highly compressed IMF inside the CIR.
 By looking at Table~\ref{tab:dlon}, we see that  $\Delta\varphi_4^{0}=4.7$,  that is, the longitudinal width of the intensity distribution is more than six times smaller than the width of the injection region. 
Comparing panel A with panel B of Fig.~\ref{fig:lon_4}, we see that cross-field diffusion has brought particles to field lines that only enter the CIR at larger radial distances.  
This can be seen by noting  that cross-field diffusion has produced substantial particle intensities  to the left of the red band denoting the reverse shock, and to the right of the blue band denoting the forward shock.  
These intensities correspond to particles that have moved to field lines that are outside the CIR at 1.5 AU, since the two shocks are the boundaries of the CIR. These field lines eventually enter the CIR at larger radial distances as a consequence of the propagation of the reverse shock into the fast solar wind and the forward shock into the slow solar wind (see also Fig.~6 of Paper~I).

Among the four cases studied, case~4 is the most influenced by cross-field diffusion since the increase of the longitudinal width due to the effect of the perpendicular diffusion is the largest, $\Delta\varphi_4^{\perp} / \Delta\varphi_4^{0} = 7.3$ (see Table~\ref{tab:dlon}). That is, the longitudinal width of the particle spread is more than seven times wider than the width when only parallel transport is considered. 
Similar to case~2, this can again be attributed to small cross-field motions inside the CIR, bringing particles to IMF lines that are closely compressed inside the CIR but significantly separated outside the CIR. 

Furthermore, Fig.~\ref{fig:lon_4} shows that the case without cross-field diffusion (panel A) results in an accelerated particle population only at the reverse shock, whereas the case with cross-field diffusion (panel B) shows the formation of accelerated particle populations centred on the forward shock and on the reverse shock (see also Paper~I); the SI separates both populations and exhibits a depressed level of the intensity (brown curve).
Such an intensity dip of the CIR-accelerated energetic particles situated near the SI  is a feature regularly observed  \citep[see e.g.][]{mason97, dwyer97, strauss16}. The reason for this phenomena is likely that the SI acts as a diffusion barrier, keeping the energetic particles accelerated at the forward shock separated from those accelerated at the reverse shock.

To conclude this section, we would like to remark that by defining $\Delta \varphi_i^{0,\perp}$  as the longitudinal width of the energy integrated intensities, we lost information on the energy dependence of the longitudinal widths. 
However, by looking at Figs.~\ref{fig:lon_1}--\ref{fig:lon_4} we see that different energy channels have different longitudinal widths, and that the energy channel that determines the total longitudinal width, $\Delta \varphi_i^{0,\perp}$,  changes across the different cases.
We find a similar behaviour for the variables 
%A similar observation can be made for the parameters
$\Delta \varphi_{\rm{L,U},i}$, which measure the longitudinal increase of the energy integrated intensities due to cross-field diffusion. 
A comparison of Figs.~\ref{fig:lon_1}--\ref{fig:lon_4} illustrates again that the cross-field diffusion does not affect all energy channels in the same manner, which is partly due to the dependence of the cross-field diffusion on the particle speed (see footnote~\ref{fn:vdependence} and Paper~I). 
The variation of $\Delta \varphi_i^{0,\perp}$ and    $\Delta \varphi_{\rm{L,U},i}$ with the energy is
%These energy dependences of $\Delta \varphi_i^{0,\perp}$ and    $\varphi_{\rm{L,U},i}$ are 
 also influenced by our choice for a mono-energetic impulsive injection. 
For example, in our simulations, the evolution of the longitudinal width of the $2.6 \pm 0.2$~MeV energy channel depends on how fast and where this channel gets significantly populated through adiabatic deceleration; thus this depends on the specific solar wind conditions that are different for the four cases. 
If we were to instead consider a  mono-energetic  delta injection of, for example 3 MeV protons instead of 4 MeV protons, the spreading of the protons populating the $2.6 \pm 0.2$~MeV  energy channel would be altered as compared to our current simulations. This is because this energy channel would now be populated from the very start of the simulation and be depleted from particles through adiabatic deceleration. 
Hence, in order to further investigate %to get a full view on 
the energy dependence of the particle spreading in the heliosphere, we need to consider a particle source injection with an energy spectrum %n energy injection spectrum
covering a wide range of energies. This task %something which 
will be addressed in a future work.

\subsection{Radial evolution of the longitudinal width}
\begin{figure}
        \centering
        \begin{tabular}{c}
        \includegraphics[width=0.4\textwidth]{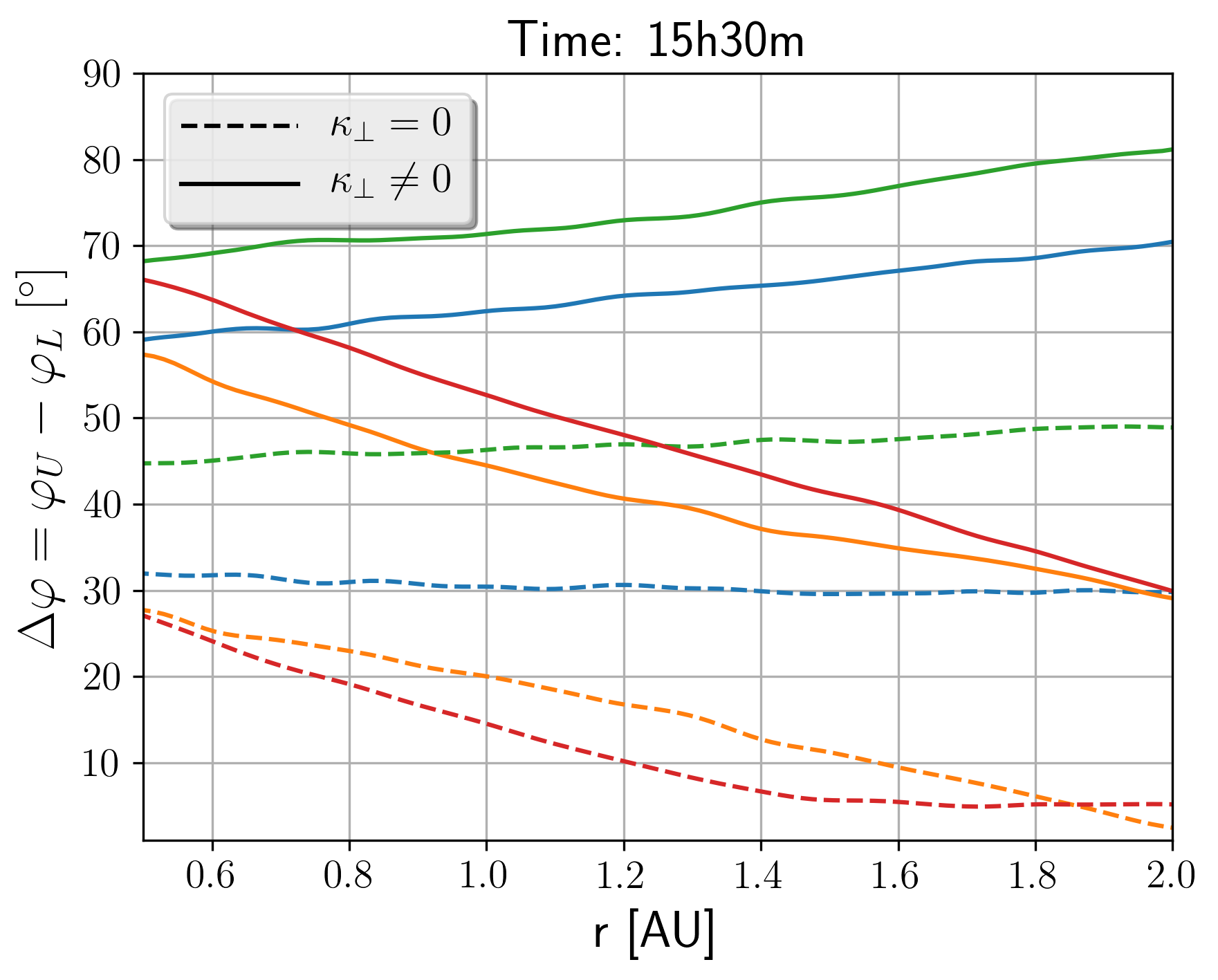}\\
        \includegraphics[width=0.4\textwidth]{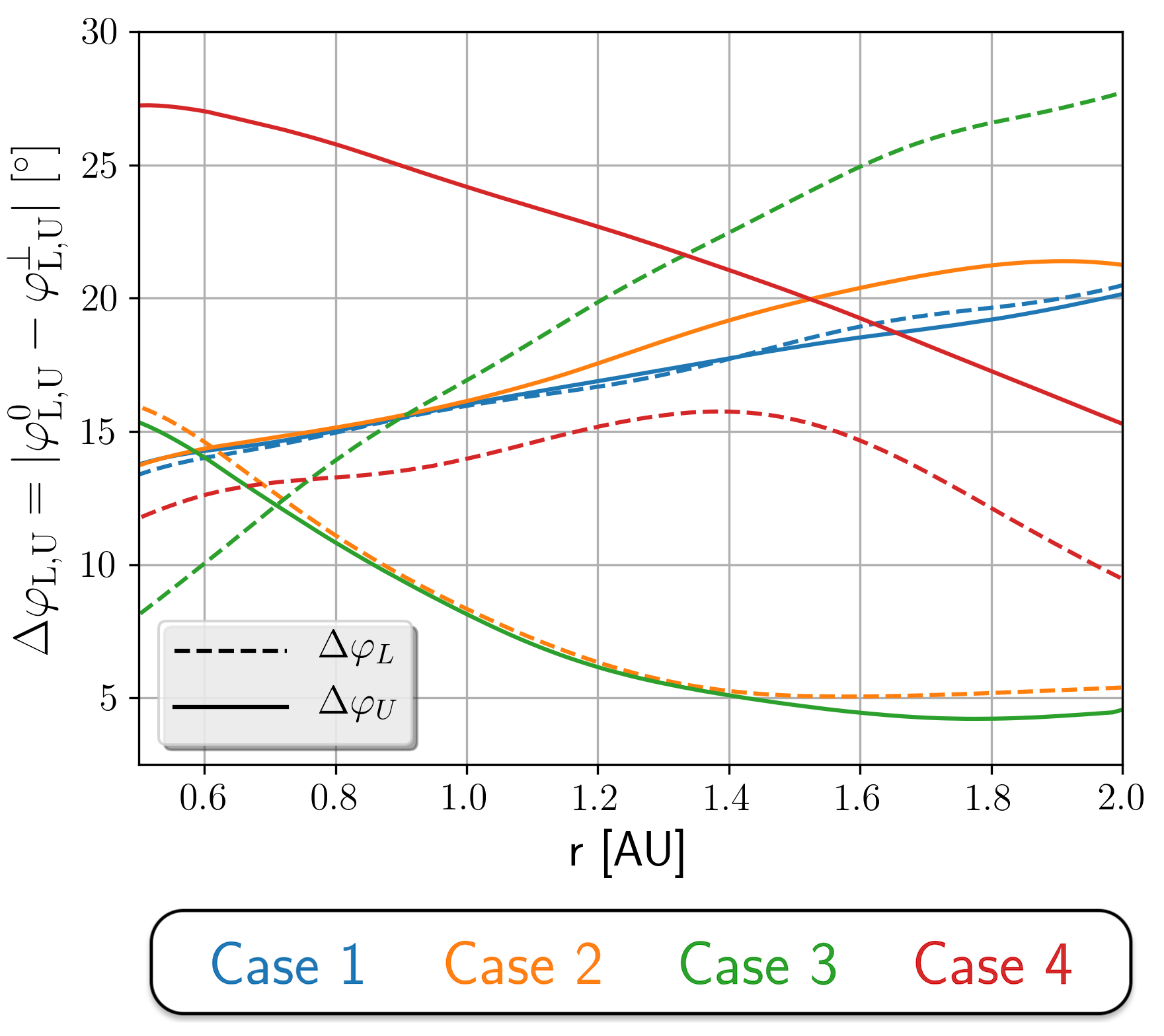}
    \end{tabular}
    \caption{\textit{Upper panel:} longitudinal width of the particle streaming zone for the simulations without (dashed lines) and with (solid lines) cross-field diffusion, 15.5 hours after particle injection. Blue corresponds to case~1, orange to case~2, green to case~3, and green to case~4.
    \textit{Lower panel:} the longitudinal increase of the particle streaming zone at the lower (dashed) and upper (solid) longitude boundaries,  15.5 hours after particle injection. The colour code follows the upper panel.  }
     \label{fig:lon_width}
\end{figure}
In the previous section, we examined how the total longitudinal width $\Delta\varphi_i$ of the particle streaming zone at 1.5 AU strongly varies across the four different cases, reflecting the underlying IMF topology.
We now study how $\Delta\varphi_i$ varies as a function of radial distance in the solar equatorial plane, 15.5 hours after particle injection. 
%More precisely, we define $\Delta\varphi_i$ for all cases as the longitudinal width of the region where the omni-directional intensity, integrated over all energies, is larger than $10^{-9}$. 
The radial dependence of $\Delta\varphi_i$ is illustrated in the  upper panel of Fig.~\ref{fig:lon_width}, for the four injection cases, both without (dashed curves) and with (solid curves) cross-field diffusion.
%From the previous section we know that the increase in the width of the particle streaming zone due  to cross-field diffusion can be highly asymmetric. Therefore, denoting the lower and upper longitudinal  edges  of the particle streaming zone by, respectively, $\varphi_{\rm{L},i}$ and $\varphi_{\rm{U},i}$, we measure the width increase at both sides of the streaming zone as $\Delta\varphi_{\rm{L},i} = \varphi_{\rm{L},i}^{0} - \varphi_{\rm{L},i}^{\perp}$ and 
%$\Delta\varphi_{\rm{U},i} = \varphi_{\rm{U},i}^{\perp} - \varphi_{\rm{U},i}^{0}$, where, like in the previous section, the superscripts $0$ and ${\perp}$ refer to the simulations without and with cross-field diffusion, respectively. 
The lower panel of Fig.~\ref{fig:lon_width} shows the radial dependence of $\Delta\varphi_{\rm{L},i}$ (dashed curves) and $\Delta\varphi_{\rm{U},i}$ (solid curves) for the four different cases. 

For case~1 (blue lines), we see that $\Delta\varphi_1^{0}$ is constant and equal to $~30^\circ$, i.e. the width of the injection zone, which is expected for a nominal IMF. Instead, cross-field diffusion causes $\Delta\varphi_1^{\perp}$ increase with the heliocentric radial distance 
%The cross-field diffusion increases  $\Delta\varphi_1$  with increasing radial distance, which is also expected 
since the cross-field diffusion in our simulations, as in others \citep[e.g.][]{Zhang09,droge10,droge14}, is stronger at larger radial distances because of the decreasing magnetic field. 
From the lower panel of Fig.~\ref{fig:lon_width}, we see that the increase in width of the streaming zone is the same at both sides of the streaming zone. 
For case~2, the evolution of $\Delta\varphi_2$  is illustrated by  the orange lines in  Fig.~\ref{fig:lon_width}. 
 In contrast to case 1, the longitudinal width of the particle intensity distribution is monotonically decreasing with radial distance in the heliospheric equatorial plane, both for the simulations with and without cross-field diffusion. 
This is because the magnetic field lines bounding injection region 2 enter the CIR, where they are compressed and, once inside the CIR, the field lines further converge towards the SI therefore decreasing the width of the streaming zone monotonically.
 Looking at the lower panel of Fig.~\ref{fig:lon_width}, we see that there is a strong difference between the change of the width at the lower and upper longitudinal borders of the streaming zone. 
 The field line bounding  the lower longitudinal side of the injection region enters the CIR at $\sim 1.3$ AU, i.e. around the location where $\Delta\varphi_{\rm{L},i}$ becomes constant and equal to $\sim 5^\circ$. 
 
%In the shocked slow wind, the IMF lines converge towards the SI, i.e., towards a stronger magnetic field where the cross-field diffusion is impeded. 
%Similarly, the field line bounding  the upper longitudinal side of the injection region enters the CIR at $\sim 1.9$ AU, i.e., around the location where $\Delta\varphi_{\rm{U},i}$ becomes constant and equal to $\sim 21^\circ$. 

As noted before, any cross-field transport inside the CIR can bring the particles on IMF lines that are closely adjacent in the CIR, but significantly separated outside the CIR. Hence, we expect to see the effect of cross-field diffusion happening inside the CIR more clearly outside the CIR, that is, before the IMF lines enter the CIR.
Cross-field transport at the lower longitudinal border in the direction of the SI are only visible at small radial distances, since the particles move on field lines that  are  everywhere closely adjacent to each other except at very small radial distances when they reside in the unperturbed solar wind.  
This explains why $\Delta\varphi_{\rm{L},2}$  is slightly larger than $\Delta\varphi_{\rm{L},1}$ below 0.6 AU.  For the same reason, $\Delta\varphi_{\rm{U},2}> \Delta\varphi_{\rm{U},1}$, since  cross-field transport  of particles inside the CIR near the upper longitude border and in the opposite direction of the SI brings particles on field lines that enter the CIR at larger radial distances. 
We want to remark that in this latter case the effect of the cross-field transport needs time to be particularly notable, since particles have to travel to large radial distances, diffuse across the magnetic field, and subsequently return towards the inner heliosphere. For this reason $\Delta\varphi_{\rm{U},2} \approx \Delta\varphi_{\rm{U},1}$ for $r<1$ AU.

For Case~3, $\Delta\varphi_3^{0}$ is slowly increasing instead of remaining constant like case~1 or decreasing like case~2.  Most of the IMF lines originating from injection region 3 enter the CIR at large radial distances and hence undergo compression at the reverse shock. However some of the IMF lines have enter the rarefaction region behind the fast solar wind stream, where IMF lines are strongly diverging. 
%This is despite the fact that most IMF lines originating  from injection region~3 also eventually enter the CIR where they are compressed in the reverse shock. However, some IMF lines enter the rarefaction region, behind the fast solar wind stream, where the field lines are strongly diverging. 
This explains the larger longitudinal width $\Delta\varphi_3$ of this case as compared to the other cases. Such large extent of the particles is also clearly observed by comparing the particle density map of the panels in the last
%something that is also clearly visible from the last  
 column of Fig.~\ref{fig:field_lines}.
The behaviours of $\Delta\varphi_{\rm{L},3}$ and $\Delta\varphi_{\rm{U},3}$ are analogous yet mirrored when compared to the corresponding variables of case~2, since the upper longitudinal edge is now close to the SI. In the lower edge, 
we note that $\Delta\varphi_{\rm{L},3} > \Delta\varphi_{\rm{L},1}$ 
for $r>1$ AU, indicating that cross-field diffusion is more efficient in the rarefaction region than in the slow solar wind because of the lower magnetic field strength in the rarefaction region that translates into a stronger cross-field diffusion (see Eq.~\eqref{eq:cross_field}).
%something that can be attributed to the lower magnetic field magnitude and hence stronger cross-field diffusion (see Eq.~\eqref{eq:cross_field}) in the rarefaction region.}

%Finally, we note that the flattening of $\Delta\varphi_3^\perp$  and $\Delta\varphi_{\rm{U},3}$ for $r\grtsim 1.8$ AU is because the intensities drop below our prescribed threshold, since the particle density is diluted over such a wide region.   
  
For Case~4, we see that $\Delta\varphi_4^{0}$ and $\Delta\varphi_4^{\perp}$ both decrease, similar to case~2. The behaviour of $\Delta\varphi_4^{0}$ is largely determined by lower longitudinal edge, since the field line bounding the  upper longitudinal edge quickly enters the CIR and converges towards the SI. Around $\sim 1.4$ AU, the field line bounding the lower longitudinal edge enters the CIR, explaining the kink in $\Delta\varphi_4^{0}$ at this distance. In contrast to the IMF in the shocked slow solar wind, the IMF in the shocked fast solar converges only very slowly in the direction of the SI, explaining why $\Delta\varphi_4^{0}$ remains approximately constant for $r > 1.4$ AU.    Figure~\ref{fig:lon_width} also illustrates that the strong increase  of $\Delta\varphi_4^{\perp}$ with decreasing radial distance is largely determined by $\Delta\varphi_{\rm{U},4}$, i.e. by cross-field motions of particles in the shocked slow solar wind. As for case~2, these motions bring particles on IMF lines that enter the CIR from the slow solar wind at much larger radial distances than the IMF lines bounding the particle streaming zone in the case with null cross-field diffusion (see the red solid line in the lower panel of Fig.~\ref{fig:lon_width}). 
%when there is no cross-field diffusion. 
The large curvature of the IMF in the slow solar wind then explains the strong increase in $\Delta\varphi_{\rm{U},4}$ towards the Sun. 
\begin{figure*}
        \centering
        \begin{tabular}{cc}
        \includegraphics[width=0.45\textwidth]{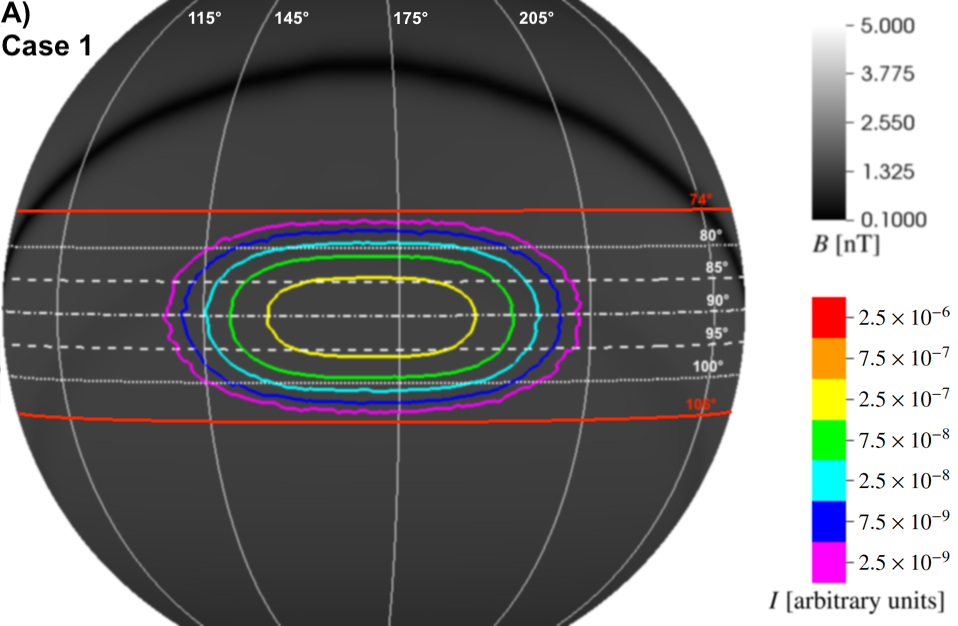}&
        \includegraphics[width=0.45\textwidth]{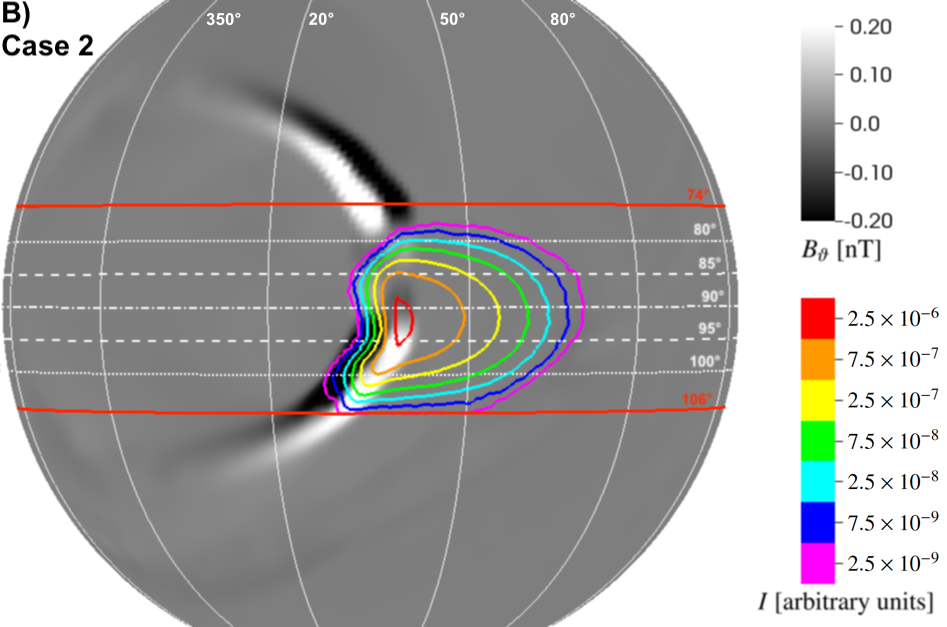}\\        \includegraphics[width=0.45\textwidth]{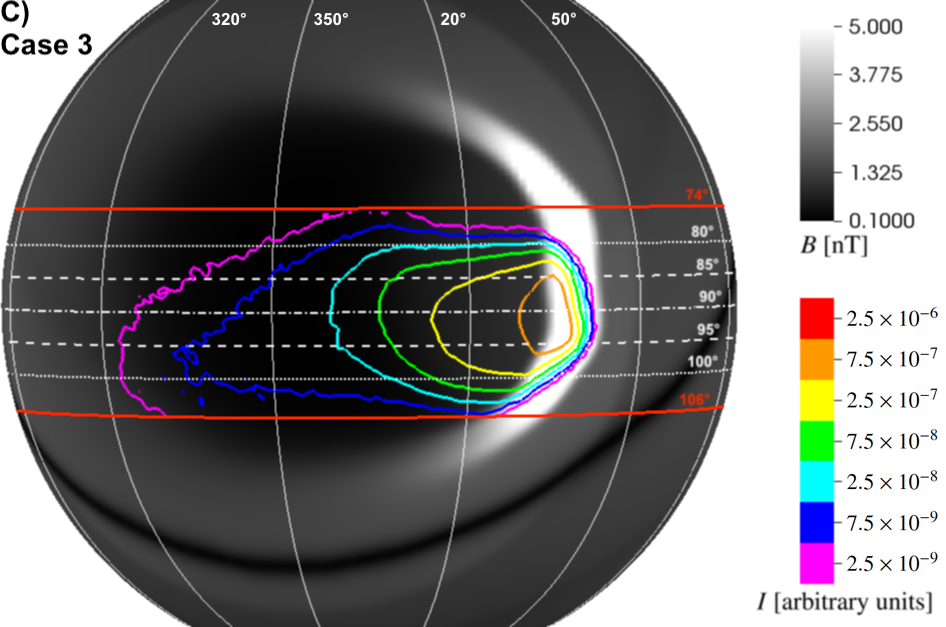}&
        \includegraphics[width=0.45\textwidth]{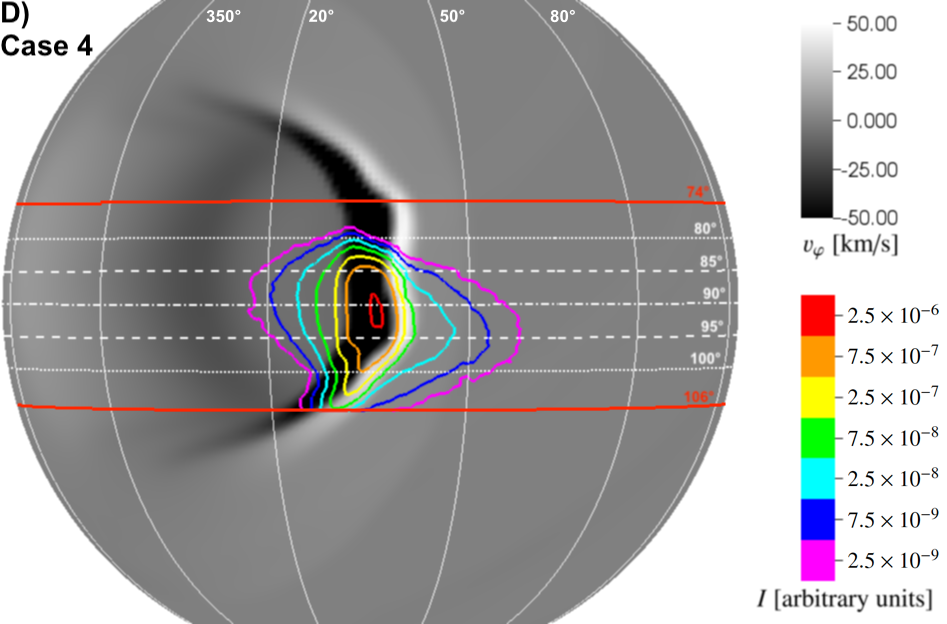}\\
        
    \end{tabular}
    \caption{Contour plots of the particle intensity at $r=1.5$ AU, drawn on top of different MHD solar wind variables, 15.5 hours after particle injection and for the simulations with cross-field diffusion. The red parallels indicate the borders of the sampling region. \textit{Upper left panel:} intensities of case~1 drawn on top of the magnetic field magnitude. 
    \textit{Upper right panel:} intensities of case~2 drawn on top of the  magnetic field  colatitude component.  
    \textit{Lower left panel:} intensities of case~3 drawn on top of the magnetic field magnitude.
     \textit{Lower right panel:} intensities of case~4 drawn on top of the longitudinal velocity component.}
     \label{fig:2D_clt}
\end{figure*}

\subsection{Latitudinal variation of particle intensity}
In the following, we examine the energy-integrated particle intensity variations both in longitude and colatitude, at a fixed heliospheric radial distance of $r=1.5$ AU and 15.5 hours after particle injection. Figure~\ref{fig:2D_clt} shows such spherical surfaces for the four different cases with the SEP intensity levels depicted as contour lines on top of different solar wind variables. All particle intensity contours represent the simulations with cross-field diffusion.
  
Panel~A of Fig.~\ref{fig:2D_clt} corresponds to case~1, where the elliptical shape of the intensity contours reflects the rectangular shape of the particle injection region (see Fig.~\ref{fig:injection_regions}). 
Although not shown in this figure, the intensity contours for the simulation with zero cross-field diffusion, are all  coincident and rectangular like the source region, as expected in a Parker spiral magnetic field.
In contrast, the shapes of the intensity contours in case~2 (panel~B Fig.~\ref{fig:2D_clt}) are no longer similar to the original source region. 
Inside the CIR, particles were transported to the south owing to the deflection of the IMF upon crossing the forward shock, as indicated by the magnetic field  colatitude component $B_\vartheta$ drawn in the background \citep[see also e.g.][]{pizzo91}. 
Panel~C of Fig.~\ref{fig:2D_clt} shows the intensity contours for case~3, drawn on top of the magnetic field strength. 
We see that the particles are spread in a large region in the rarefaction zone behind the CIR, characterised by a weak magnetic field. 
The peak of the intensities occurs at the reverse shock and the corresponding downstream region, where the magnetic field is compressed. 
Finally, the intensity contours for case~4 (panel~D Fig.~\ref{fig:2D_clt}) are drawn on top of the longitudinal solar wind velocity component $v_\varphi$, clearly indicating the location of the SI. Similar to case~2, we see that the intensity contours are peaking towards the south in the shocked slow solar wind, as a result of the  deflection of the IMF at the forward shock. 
We note that we do not see the particle population moving towards the northern hemisphere owing to the northward location of the coronal hole
centre with respect to location of the injection region (see the left panel of Fig.~\ref{fig:injection_regions}).
The deformation of the particle intensity contours
%streaming zone
both in longitude and latitude makes  the original shape of the injection region no longer easily discernible unless the IMF structure is known accurately. 

\section{Summary}\label{sec:conclusion}

In this article we continued the work of Paper~I, studying SEP transport in a solar wind generated by EUHFORIA containing a CIR. 
We considered an impulsive injection of 4 MeV protons, originating from four source regions, all located at different places at the inner boundary of EUHFORIA. 
When examining the particle intensities in the solar equatorial plane, the four cases differed significantly from one another, reflecting the different underlying IMF structures encountered by the particles. 
Measuring the intensity and anisotropy along four different magnetic field lines illustrated the important effect of the solar wind speed on changing the energy content of the initial injected particle population.

The three IMF lines that were connected to the CIR showed significant differences from the IMF line located entirely in the slow solar wind. The intensity and anisotropy profiles along the IMF lines crossing the forward and reverse shocks demonstrated how accelerated particle populations formed centred on the shock waves because of the  presence of converging plasma flows.
The longitudinal intensity and anisotropy profiles at $r=1.5$ AU illustrated how the solar wind configuration completely alters the observed profiles. 
Whereas the longitudinal profile in a nominal solar wind is symmetric, both without and with cross-field diffusion, the longitudinal profiles measured near the CIR were highly asymmetric. 
In the latter cases, the intensities  reached their peak values at the shocks bounding the CIR, reflecting the compressed magnetic flux tubes there. 
For cases~2 and~3 we saw a sharp drop in intensities near the SI, as the corresponding injection regions were not crossing the transition region between the fast and slow solar wind at the inner boundary. 
Cross-field diffusion smoothed this drop only slightly, since the SI is characterised by a strong IMF and hence a weak cross-field diffusion. 
The effect of the SI on the particle intensities was also clearly reflected in the $4.2\pm0.2$ MeV energy channel of case~4 with cross-field diffusion. 
This energy channel is populated by two proton populations accelerated in the CIR shocks, and the SI separates both populations. Although the effect of the SI has been predicted by simulations accounting for cross-field diffusion processes \citep{strauss16}, for the first time our model is able to show with detail the acceleration of protons at both shocks of the CIR  and reproduce the observed dip of particle intensities \citep[e.g.][]{dwyer97}.

Studying the width of the particle streaming zone as a function of radial distance revealed that the different solar wind configurations produce completely different dependencies. 
For a nominal solar wind without cross-field diffusion, the width remains constant whereas in non-nominal conditions, the width can either increase or decrease with heliographic radial distance.  
Moreover, whereas the cross-field diffusion widens the particle streaming zone, it does not significantly influence the behaviour of the width as a function of radial distance.   
We suggest therefore that the IMF topology might play an important role in contributing to the occurrence of widespread and narrow SEP events. In addition we analysed the latitudinal spreading of the particles, by showing the particle distributions on a spherical surface at a heliocentric radial distance of $1.5$~AU. We show %Finally, we looked at a spherical cross-section of the particle streaming zone at a heliographic radial distance of $r=1.5$ AU. This 
that the IMF structure of the CIR alters the shape of the intensity distribution, such that the shape of the original injection region is no longer obvious.

Finally, we want to point out that large SEP events simultaneously observed by e.g., the Solar Terrestrial Relations Observatory (STEREO) and by near-Earth spacecraft might develop under different background solar wind conditions (on top of observed CME propagation), as detected by each spacecraft \citep[e.g. see the events analysed by ][]{lario14,lario16}. 
The modelling presented in this work suggests that a non-nominal structured solar wind may significantly affect both the spatial extent and energy of the particle population in SEP events. 
How the SEP spatial distribution is  affected by the solar wind also depends on the relative location of the particle source with respect to the large-scale structure. 
Therefore, models such as ours may contribute towards gaining insights in understanding the observed multi-spacecraft SEP event intensities.
%located at differ
%\alert{add a few sentences of how your modeling approach seems very competent on shedding light on the dilemma of wide spread events etc..}
%\textcolor{violet}{I want to check a couple of references more to ornament the conclusions with stronger statements on the originality of this work, }
\begin{acknowledgements}
N. Wijsen is supported by a PhD Fellowship of the Research Foundation Flanders (FWO). The computational resources and services used in this work were provided by the VSC (Flemish Supercomputer Center), funded by the Research Foundation Flanders (FWO) and the Flemish Government, department EWI. The work at KU Leuven was done in the framework of the projects GOA/2015-014 (KU Leuven), G.0A23.16N (FWO-Vlaanderen) and C 90347 (Prodex). The work at University of Barcelona was partly supported by the Spanish Ministry of Economy, Industry and Competitivity under the project AYA2016-77939-P, funded by the European Union's European Regional Development Fund (ERDF), and under the project MDM-2014-0369 of ICCUB (Unidad de Excelencia `Mar\'ia de Maeztu'). The work at University of Helsinki was carried out in the Finnish Centre of Excellence in Research of Sustainable Space (Academy of Finland grant numbers 312390 and 312351). The authors thank the referee for the helpful suggestions. We also thank Blai Sanahuja and Rami Vainio for valuable discussions.
\end{acknowledgements}

%-------------------------------------------------------------------
\bibliographystyle{aa.bst}
%\bibliography{sep_bib_AA}

\end{document}